# Exact off-resonance near fields of small-size extended hemielliptic 2-D lenses illuminated by plane waves

Artem V. Boriskin, [1,*] Ronan Sauleau, [2] and Alexander I. Nosich [1]

[1] *Institute of Radiophysics and Electronics NASU, Kharkiv, Ukraine*
[2] *Institut d'Electronique et de Télécommunications de Rennes, Université de Rennes 1, Rennes cedex, France.*
[*] *Corresponding author: a_boriskin@yahoo.com*

The near fields of small-size extended hemielliptic lenses made of rexolite and isotropic quartz and illuminated by *E*- and *H*-polarized plane waves are studied. Variations in the focal domain size, shape, and location are presented versus the angle of incidence of the incoming wave. The problem is solved numerically in a two-dimensional formulation. The accuracy of results is guaranteed by using a highly efficient numerical algorithm based on the combination of the Muller boundary integral equations, the method of analytical regularization, and the trigonometric Galerkin discretization scheme. The analysis fully accounts for the finite size of the lens as well as its curvature and thus can be considered as a reference solution for other electromagnetic solvers. Moreover, the trusted description of the focusing ability of a finite-size hemielliptic lens can be useful in the design of antenna receivers.

**1. Introduction**

Compact hemielliptic dielectric lenses made of low-index materials are used in many millimeter (mm) and sub-mm wave antennas to improve the radiation characteristics of primary feeds or, reciprocally, to enhance sensitivity of detectors [1, 2]. The elliptical shape of the front of the lens that is commonly used in such antennas is suggested by ray-optics considerations. Such a shape is capable of collecting all the rays impinging on the lens surface along the major axis in the rear focus, if the ellipse eccentricity is related to material permittivity as $e = \varepsilon^{-1/2}$. This rule generally works out well for hemielliptic lenses of comparatively large electrical size. For lenses whose size is comparable to the wavelength, the focusing ability becomes questionable. As is known, in this case there is no longer a focal point; instead, a focal spot appears whose size, shape, and location depend on the lens parameters as well as the polarization and angle of incidence of the incoming wave [3, 4]. This phenomenon is explained by the greater impact of internal reflections on the electromagnetic behavior in reduced-size lenses. It escapes accurate description in the high-frequency approximations commonly used in the analysis of dielectric lens antennas (DLA) but a better understanding of the focusing ability of small-size lenses would facilitate the design of DLAs with improved performance. For instance, such analysis can help estimate the maximum size of the receiver or feed array and determine the optimal locations for its elements.

Any dielectric lens has a finite closed boundary and therefore is, in fact, an open dielectric resonator of a certain shape. Thus its electromagnetic behavior is determined by the interplay of two major optical mechanisms, namely, geometrical-optics focusing and wavelength-scale internal resonances. The latter can be observed if the incident field frequency hits the real part of the complex-valued frequency of a natural mode of the resonator. The presence of high-Q resonances ($10^3$ and higher) depends on the lens shape, size, and dielectric constant. Intrinsic to dielectric lenses typically used in the design of mm and sub-mm wave DLAs are, for instance, whispering-gallery modes in circular and spherical lenses [5] and half-bow-tie modes in hemielliptic lenses [6]. The latter modes are closely related to the bow-tie modes in full elliptic and stadium-shape dielectric cavities studied numerically in [7]. Outside the high-Q resonances, geometrical-optics focusing is dominant. For instance, low-index ($\varepsilon \leq 4$) extended hemielliptic lenses of several wavelengths in size, typically used in integrated DLAs, generally demonstrate the priority of the focusing mechanism over the resonance one. However, the latter can become dominant for the same lenses made of high-index material ($\varepsilon \geq 10$) such as silicon [6]. This coexistence and overlap of the optical and modal features in the behaviors of small-size dielectric lenses make accurate description of their electromagnetic properties a challenging task. As has been recently demonstrated, the high-frequency approximations such as geometrical and physical optics (GO, PO) cannot fulfill this task [8] even though they are often successfully used for the analysis of DLAs, especially in the emission regime, e.g. [9-11]. This is because they fail to account for the finite curvature of the lens and hence to accurately reproduce the multiple internal reflections even for off-resonance excitation. Furthermore, accurate description of the resonance properties of such lenses is troublesome even for full-wave approaches such as FDTD [12,13].

In general, efficient focusing is a useful feature of a lens, while resonances can be viewed as undesirable. In this paper we study compact extended hemielliptic lenses made of low-index materials (rexolite and isotropic quartz) that are typical for DLAs operating in the mm-wave range and illuminated with monochromatic plane waves. The method used allows accurate characterization of electromagnetic fields both in and off the so-called half-bowtie resonances whose features have been studied in [6,8,12]; however, in contrast to the mentioned papers the goal of the present study is the off-resonance behavior of the lens. Here, we deal with the 2-D lens model. In comparison with more realistic 3-D descriptions, such a model has obvious computational advantages while retaining the capability of studying all relevant physical phenomena. Besides, the 2-D model is known to be a good approximation for a thin planar lens, provided that the bulk refractive index is replaced with its effective value.

The paper is organized as follows. The problem formulation and a brief description of the method used are given in Section 2. Numerical results are presented in Section 3, and conclusions are summarized in Section 4.

## 2. Problem formulation and outline of solution

We consider the 2D lens as a homogeneous dielectric cylinder whose profile is described by an analytical curve that is a smooth junction of a hemiellipse and a hemi-superellipse (rectangle with rounded corners [8]). Two curves are joint at the points $(0, \pm a)$, where $a$ is the minor semi-axis of the ellipse (Fig. 1). Although all parameters can be arbitrary, in computations the eccentricity of the hemielliptic frontal part is chosen in accordance with the GO focusing rule, i.e. as the inverse of the refractive index. The extension of the lens rear part is taken equal to the ellipse focal distance. For numerical simulations we select the lens material to be lossless rexolite ($\varepsilon = 2.53$) and quartz ($\varepsilon = 3.8$). The latter is assumed to be isotropic (fused silica). Their bottom size $2a$ equals $4\lambda_0$, where $\lambda_0$ is the wavelength of the incident wave in free space.

To characterize the true electromagnetic performance of finite-size lenses, we use an in-house algorithm based on the Muller boundary integral equations (BIE) and capable of providing a controlled accuracy of the numerical solution for any set of lens parameters [8,12-14]. The efficiency of the algorithm is enhanced by the method of analytical regularization and the trigonometric Galerkin discretization scheme. This leads to the matrix equation of the Fredholm second-kind type whose elements are computed as Fourier expansion coefficients of smooth or integrable BIE-kernel functions. This technique guarantees uniqueness of the solution as well as its fast monotonic convergence. Details of the mathematical approach can be found in [14] and therefore are not described here.

It must be noted that the other types of BIEs met in the literature suffer from a serious demerit: they possess an infinite number of spurious purely real eigenfrequencies [15]. They are the eigenvalues of the interior electromagnetic problem for the perfectly electrically conducting boundary filled with the material of the outer medium (usually air). Because of their presence the corresponding BIEs have largely academic interest, as in the numerical solution of the scattering problem they lead to erroneous spikes whose width depends on the coarseness of the algorithm. The frequency dependences display a "forest" of such spikes if the size of the scatterer exceeds a few wavelengths, and therefore such BIEs can be used at best in the computations of sub-wavelength scatterers (such as in [16]). In the eigenmode analysis, such BIEs lead to stable algorithms only if the eigenfrequency has a sizable imaginary part (low-Q resonances) [17,18], and they fail for high-Q ones such as whispering-gallery modes (this is admitted in [17]).

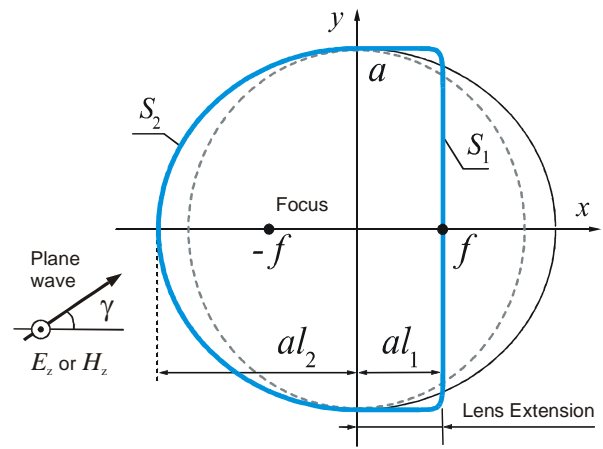

Fig. 1. Geometry and notations of the 2D model of the extended hemielliptic lens.

Fortunately, the Muller BIEs (two coupled equations in 2-D and four in 3-D) are free of the mentioned spurious eigenvalues. Being the Fredholm second-kind IEs, they can be solved numerically with virtually any nonpathological discretization scheme. This can be meshing the boundary (collocations or "pulse-basis-delta-testing" in the moment-method terminology) as in [19], Galerkin projection of the global functions [14], or a Nystrom-type interpolation scheme as in [20]. In all cases the convergence is guaranteed if the contour is a continuous and smooth curve (i.e. has a continuous derivative). Note that the size of the matrix that must be inverted is determined (see [14]) by three factors: optical size of dielectric scatterer, peak normalized curvature of its contour, and the number of desired correct digits. Thus, a blind use of the famous "rule of thumb" of taking ten points per lambda is far from being correct. More detailed comparative analyses of various BIE and also volume IE formulations for dielectric scatterers can be found in a recent review [21].

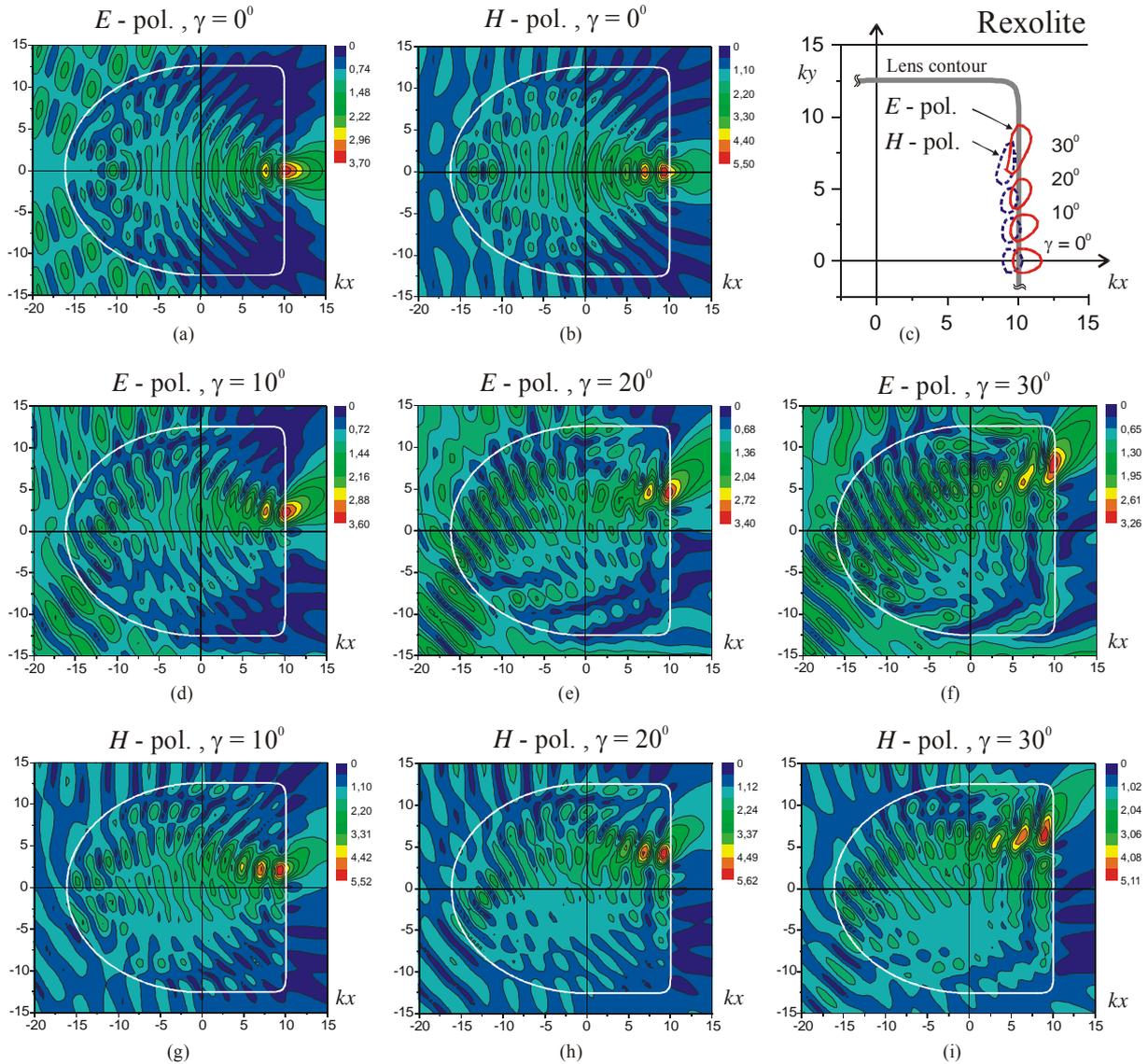

Fig. 2. (Color online) Near-field maps of rexolite extended hemielliptic lens ($\varepsilon$ =2.53, $ka$ =12.56, $l_1$ = 0.8, $l_2$ = 1.285) illuminated by unit-amplitude *E*- and *H*- polarized plane waves.

## 3. Numerical results

The near-field maps presented in Fig. 2 are for a rexolite lens illuminated by unit-amplitude *E*- and *H*-polarized plane waves impinging on the lens under various angles of incidence, $\gamma$. In the figures, the scales correspond to coordinates multiplied by the free space wavenumber, $k = 2\pi/\lambda_0$. The peak field values for each near-field map are indicated as maxima of the relevant color bars. The scale of each bar is defined as one-tenth of its maximum. For convenience of comparison, the focal spot coordinates obtained for various angles of incidence are put together in Fig. 2c. Here, the focal spot's boundary corresponds to 80% of the peak field amplitude value (or 64% of the peak field intensity value). The term "focal spot" means the spot containing the highest field value and located closest to the focal point predicted by GO.

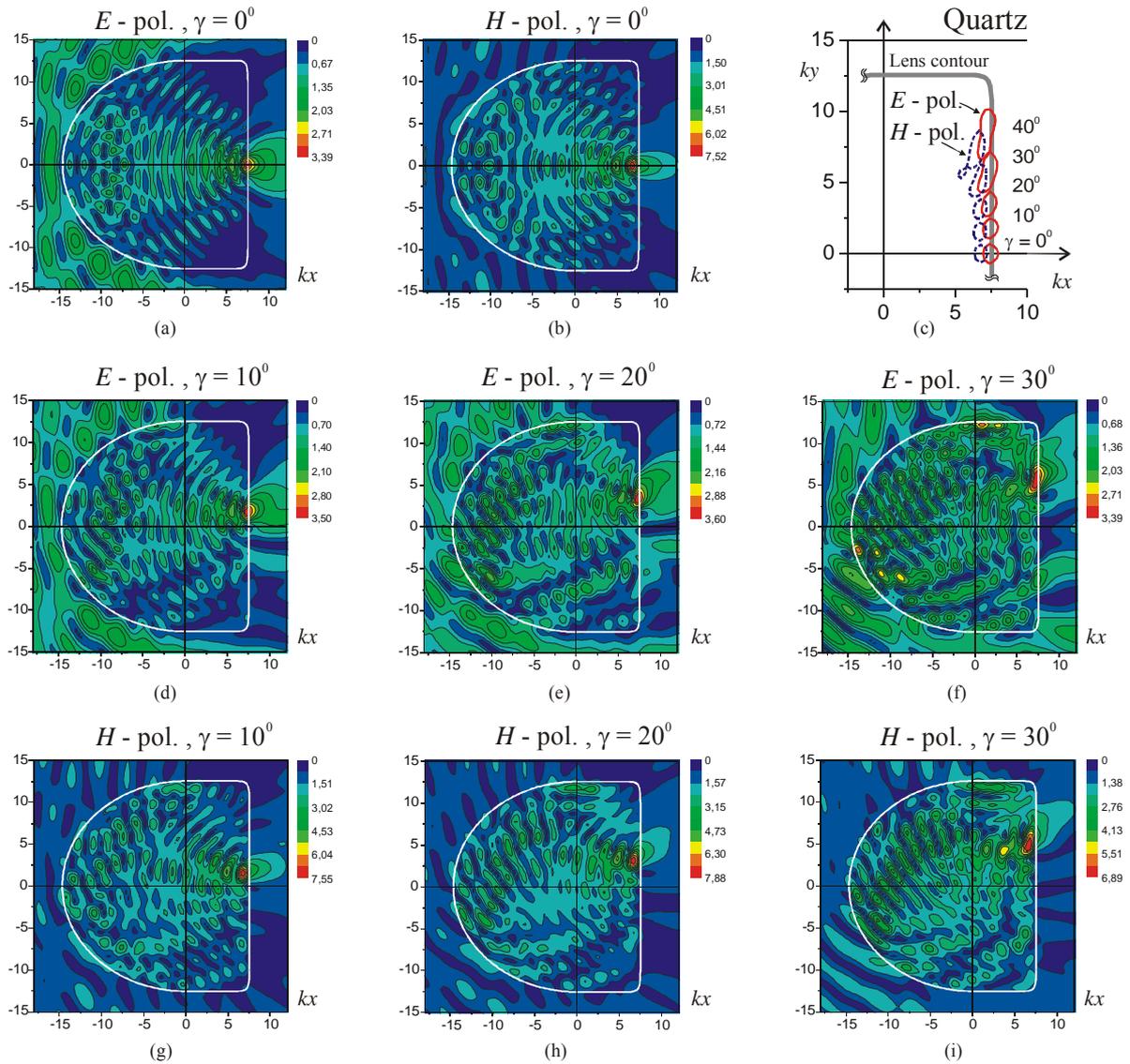

Fig. 3. (Color online) Near-field maps of quartz extended hemielliptic lens ($\varepsilon = 3.8$, $ka = 12.56$, $l_1 = 0.6$, $l_2 = 1.165$) illuminated by unit-amplitude *E*- and *H*- polarized plane waves.

Note that for the larger angles of incidence several spots with comparable values of the field appear (see Fig. 3f).

The field patterns within the lens clearly show the importance of internal reflections. In contrast to GO, which predicts a single focal point in the ellipse rear focus, one can see a standing wave pattern with several spots of almost equal field amplitudes. For the varying illumination angle ($\gamma \neq 0°$) these spots migrate along the flat bottom of the lens and vary both in size and shape. Moreover, the spots size, shape, and location also depend on the polarization; that is in the *H*-polarization the focal spots of the $H_z$ component are shifted by $\lambda/4$ inside the lens and their normalized peak field amplitudes are about 1.5 times higher than those for the *E*-polarization. The former is observed because of the phase shift of $\pi/4$ that exists between the *E* and *H* field components, whereas the latter is explained by the difference in the boundary conditions. Namely, all field components are continuous across the air-dielectric boundary in the *E*-case, while a jump in the *E*-field normal component takes place in the *H*-case, proportional to $\varepsilon$. Finally, the difference in size and shape of the spots observed for the *E*- and *H*-polarizations appears due to the difference in the reflection/transmission properties for the *E*- and *H*-waves propagating inside the lens and affecting the focal spot formation (see Appendix A).

The near fields of the isotropic-quartz lens illuminated by the plane *E*- and *H*-waves (Fig. 3) reveal a greater role played by internal reflections in the formation of the field patterns than for the lenses made of less dense materials. The high-field spots (including the focal spots) become smaller in size evidently because the wavelength in quartz is shorter than in rexolite.

It is interesting to note that for each material the size and shape of the focal spots observed for all incidence angles $\gamma \leq 20°$ remain the same for either of the two polarizations (note the curves characterizing the focal spots' behavior as given in Fig. 4). This can be explained by the increase in the lens's receiving aperture that compensates to some extent for the loss of the lens's focusing ability due to aberrations inherent to any realistic lens. For $\gamma > 20°$ the spots become twice as large in size and the peak intensity decreases rapidly (the near-field maps for a quartz lens illuminated at 40° are omitted for brevity, but the corresponding focal spot contours for both polarizations are given in Fig. 3c). These observations indicate the maximum size of a focal array of receivers that can be used in the design of a beam-switching DLA. Moreover, the knowledge of the focal spot sizes and locations for different angles of the plane wave incidence can help determine the optimal locations and spacing for focal array elements.

In order to extend the results of the near-field analysis to lenses of other sizes, we have plotted the peak-field values defined as the highest field amplitude within the focal spot, for rexolite and quartz lenses symmetrically illuminated by plane waves, versus the normalized frequency (Fig. 5). As one can see, the peak-field value grows proportionally to the electrical size of the lens. The oscillations observed highlight the presence of internal reflections and their impact on the focal spot formation. This is confirmed by the higher amplitude of ripples observed for the quartz lens. The difference between polarizations (roughly a factor of $\varepsilon^{1/2}$) can be explained by the different level of back reflection of the incident wave and by the difference in transparency of the lens boundary for the "rays" inside the lens (see Appendix A).

As to the location of the focal spots, it remains unchanged for the lenses of any size, that is for the *E*-polarization the $E_z$-component focal spots lie at the lens boundary, whereas for the *H*-polarization the $H_z$-component spots are shifted inside the lens for about a quarter of the wavelength in the material (Fig. 6).

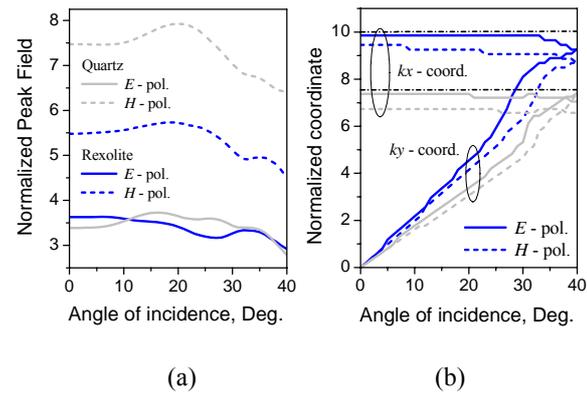

(a)          (b)

Fig. 4. (a) Normalized peak-field values in the focal spots of the rexolite and quartz lenses and (b) normalized coordinates of the corresponding points versus the angle of incidence of plane waves. Dashed-dotted lines in (b) indicate the size of the lens extension ($al_1$).

## 4. Conclusions

The previously developed highly efficient numerical algorithm based on the Muller BIE has been applied to the analysis of 2D models of compact extended hemielliptic dielectric lenses whose size and materials are typical for mm-wave lens antennas. The exact near fields for the lenses excited by the plane waves under various angles of incidence have been plotted. They demonstrate nontrivial electromagnetic behavior of the small-size lenses that cannot be accurately characterized by the conventional high-frequency methods often applied to the analysis of such lenses.

Studying off-resonance focusing by moderately small lenses, we have found that with a rexolite or quartz lens of the cross-sectional size of 3 to 5 free-space wavelengths one can obtain focal-spot peak field values that are 3 to 8 times larger than in the incident plane wave. These numerical results are accurate and thus can be considered as a reference solution for other electromagnetic solvers; they can also serve as a guideline for design of antennas and lenses.

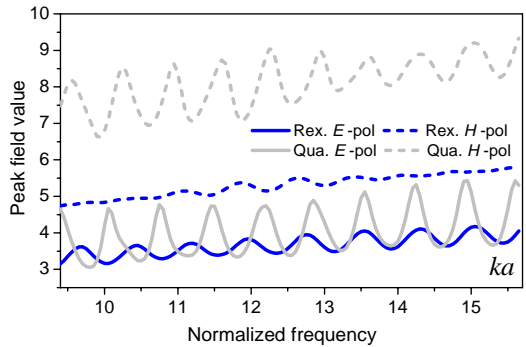

Fig. 5. Normalized peak field values in the focal spots of the rexolite and quartz lenses symmetrically ($\gamma = 0°$) illuminated by unit-amplitude plane *E*- and *H*- waves versus the normalized frequency.

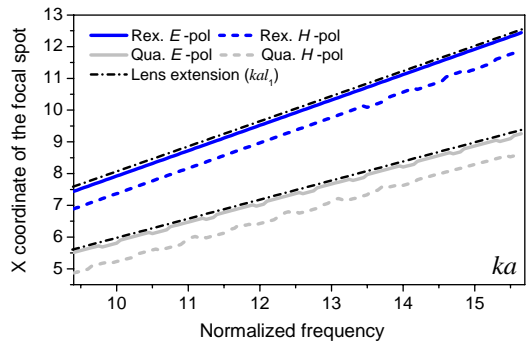

Fig. 6. Horizontal coordinates ($kx$) of the points with the highest field values defined in the focal spots of the rexolite and quarts lenses illuminated symmetrically ($\gamma = 0°$) by plane *E*- and *H*- waves versus the normalized frequency. The dashed-dotted lines indicate the normalized sizes of the lens extensions for both lenses. The field values at the corresponding points are given in Fig. 5.

## Acknowledgments

This work was supported in part by joint projects between the Institute of Radiophysics and Electronics of the National Academy of Sciences of Ukraine, on the one side, and the Centre National de la Recherche Scientifique, Ministère de l'Education Nationale, de l'Enseignement Supérieur et de la Recherche, and Ministère des Affaires Etrangères et Européennes, France, on the other side. The first author was also supported by the Foundation Michel Métivier, the Brittany Region via the CREATE/CONFOCAL project, and by the North Atlantic Treaty Organization (NATO) via grant NIKR.RIG.983313.

## Appendix A: Reflection and transmission coefficients for the plane E- and H-waves incident on the air-dielectric boundary

Transmission and reflection coefficients of a plane wave arbitrarily incident on a plane air-dielectric boundary can be determined using the classical Fresnel formulas. The reflection coefficient dependences on the angle of incidence are shown in Fig. 7 to illustrate the discussion of the numerical results obtained for the considered lenses.

Although for an extended hemielliptic lens of compact size the curvature of the profile is high and thus cannot be neglected, the curves in Fig. 7 for the plane waves impinging on the rexolite and quartz boundaries with air help us to understand the differences in the impact of internal reflections on near-field pattern formation.

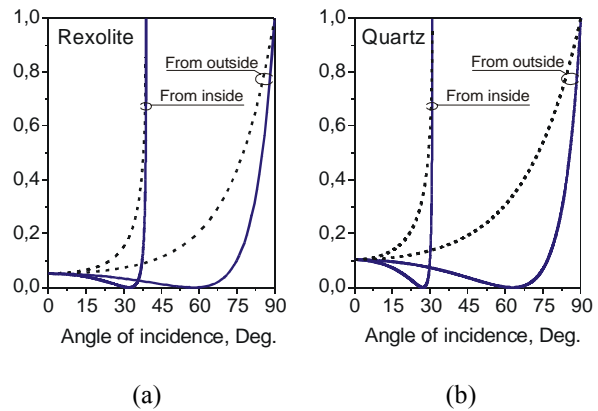

Fig. 7. Intensity reflection coefficients for the *E*- (solid curve) and *H*-polarized (dashed curve) plane waves illuminating the flat air-dielectric surface of: (a) rexolite, (b) quartz.